\documentclass{sig-alternate}

\usepackage{amssymb}
\usepackage{graphicx}

\usepackage{url}
\usepackage{booktabs}
\usepackage{dcolumn}

\usepackage{balance}  
\usepackage{graphics} 
\usepackage{times}    
\usepackage{url}      
\usepackage{amsmath}
\usepackage[usenames,dvipsnames]{xcolor}
\usepackage{enumitem} 
\usepackage{tabularx}

\newcounter{NumberOfComments}
\stepcounter{NumberOfComments}

\definecolor{DarkGreen}{rgb}{0.000000,0.6,0.000000 } 

\newcounter{JSNumberOfComments}
\stepcounter{JSNumberOfComments}

\begin{document}
%
\conferenceinfo{Computation+Jounalism Symposium}{'14 New York USA}

\title{Understanding News Geography and Major Determinants of Global News Coverage of Disasters}
%
%
%
%
%

\numberofauthors{2} 
%
\author{
%
%
\alignauthor
Haewoon Kwak\\
       \affaddr{Qatar Computing Research Institute}\\
       \affaddr{Doha, Qatar}\\
       \email{hkwak@qf.org.qa}
\alignauthor
Jisun An\\
       \affaddr{Qatar Computing Research Institute}\\
       \affaddr{Doha, Qatar}\\
       \email{jan@qf.org.qa}
}

\maketitle
\begin{abstract}
In this work, we reveal the structure of global news coverage of disasters and its determinants by using a large-scale news coverage dataset collected by the GDELT (Global Data on Events, Location, and Tone) project that monitors news media in over 100 languages from the whole world.  
Significant variables in our hierarchical (mixed-effect) regression model, such as population, political stability, damage, and more, are well aligned with a series of previous research. 
However, we find strong regionalism in news geography, highlighting the necessity of comprehensive datasets for the study of global news coverage.
\end{abstract}




\section{Introduction}

Traditional mass media have played a significant role in delivering foreign news~\cite{ball1976dependency}.  Even in the era of social media, traditional news channels are still the main sources that most people rely on~\cite{leckner2011sampler}. 
Thus, the selection of foreign news by domestic news media is shaping individuals' perception about those countries~\cite{wanta2004agenda}.

A wide news network woven by international news agencies helps to transform a remote disaster into an international crisis. Even though a majority of disasters remains unreported~\cite{franks2006carma}, the reported ones evoke compassion, and this potentially leads to various charitable acts, such as fund-raising to provide monetary support.  These days it is not uncommon to expect help from the world when a tragedic disaster happens. In this sense, global news coverage of a disaster is a sufficient condition for worldwide public action.  Then, a central question naturally arises: which disasters are covered and which are not?  The systematic approach to address this question requires a comprehensive dataset of news media sites in different countries over long period, which remains to be unexplored in traditional media research.  We revisit previous research based on a single country or a region~\cite{galtung1965structure,westerstaahl1994foreign} and examine whether it also holds globally. 

Focusing on disasters only brings some advantages in studying global news coverage.  Concern for others is one of the fundamental behavioral mechanisms of human beings, which can be observed from infancy~\cite{zahn1992development}.  Such traits explain why news about victims of disasters is chosen to be reported even when no economic or political incentives are expected.  Also, news stories on natural disasters are generally not filtered out by censorship which might distort normal news flow among countries. 

Furthermore, unlike the other type of news that comes from a few constant dominant sources (e.g., US entertainment), there are no necessities that this is the case for disaster news. The US is well known as the center of the global economy and politics, and at the same time its culture-related industries are strong enough to generate a lot of entertainment news. Consequently, a huge flux of entertainment news from the US to the rest of the world cannot be avoided. We expect such flow whenever prominent news sources exist for particular news types. Without such prominent sources, as a result, we expect that the global news coverage of disasters correctly captures the newsworthiness of events.  

In this work we use a large-scale news media coverage dataset collected by the GDELT (Global Data on Events, Location, and Tone) project. GDELT project monitors news media in over 100 languages from the whole world~\cite{leetaru2013gdelt}. With large-scale data of 195 thousand disasters happening from April 2013 to July 2014~\cite{gdelt@dataset}, we examine which disasters receive a great deal of attention from foreign news media.  

Our answer entails two elements. One is revealing the news geography, describing the general pattern for which countries are presented in which other countries' news.  This is one of the central issues regarding international news studies from the 1970s~\cite{gerbner1977many}.  We connect our findings to the well-established framework of news geography~\cite{sreberny1984world}.  
The other is investigating the major determinants of global news coverage.  This also has been studied for decades with various datasets collected in different countries, but without providing a holistic view of news coverage of disasters.  
We build a hierarchical (mixed-effect) multiple regression model whose dependent variable is the number of countries covering the disaster and whose independent variables are a wide range of disaster-, nation-, and news logistics-related attributes.  
We find that some variables known to be relevant to a news coverage of a foreign disaster from studies of a single country or a region, such as population, are significant in our model as well.  They still matter from the global view.  However, strong regionalism found in news geography posits that foreign news coverage greatly varies by counties, and highlights the importance of the comprehensive dataset for the correct understanding of global news coverage.




\clearpage
\section{Theoretical Orientation}


\subsection{Theory of newsworthiness}

Foreign news coverage is the outcome of a news selection process~\cite{chang1992factors}.  Making a decision on which news items to report is essential  because news is delivered through physically limited channels, such as pages in newspapers and minutes in TV news. 



One of the seminal studies that examines the factors of news selection was conducted by Galtung and Ruge~\cite{galtung1965structure}. They propose the theory of newsworthiness that is based on psychology of individual perception and explain which factors influence newsworthiness of an event. The suggested factors are frequency, intensity, unambiguity, meaningfulness, consonance, unexpectedness, continuity of an event, and some characteristics (e.g. identity) of an actor involved in the event.  Some critics argue that applying their theory to the global news flow between nations is insufficient owing to the lack of systematic determinants based on the power structures of the world~\cite{wu1998investigating,wu2003homogeneity}
Nevertheless, the theory of newsworthiness has provided a foundation of subsequent news flow studies with a few variations of some factors~\cite{harcup2001news}.  We consider several factors among their suggestions, such as unexpectedness and intensity of an event and the identity of an actor extracted from the GDELT dataset.  

In the extension of their study, deviance of an event is found to be an important factor~\cite{shoemaker1991deviant}.  In homicide incidents, deviance defined by the gender and the race of victims and offenders is known to be prominent factors in news selection~\cite{pritchard1985race}. We incorporate race and gender information of victims if available. 

The significance of the number of people killed by a disaster in predicting its news coverage is still debatable.  Gaddy and Tanjong reported its importance~\cite{gaddy1986earthquake}, while others found no significance~\cite{adams1986whose}.  Yet, the number of victims is a good proxy to reflect the intensity of a disaster, especially when the other measures for the damage of a disaster is unavailable.  We thus include the number of victims of a disaster in our study and validate its importance. 

\subsection{Effect of national attributes on foreign news coverage}

While determinants of global news flow, in terms of the amount and its direction, have been repeatedly  investigated based on a single or few countries for decades~\cite{ostgaard1965factors}, results are not consistent between countries, mainly due to cultural difference~\cite{peterson1979foreign}.  

One of the earliest studies reporting general patterns of global news flow was conducted with the `Foreign News' dataset that contains news media of 46 countries for two weeks in 1995~\cite{sreberny1999comparative}.  The findings are relatively stable, and confirmed by subsequent studies.  We note that these studies are usually conducted from the view of a guest and a host country relationship.  The guest country is where the event happens, and the host country is where the news media exist.  In this view, the problem of global news flow is transformed into the problem of dyadic relationship between countries, like whether an event in a certain guest country is covered in a particular host country.  The general factors to affect the news coverage of a host country can be divided into two categories.  One is the attributes of a guest country, and the other is proximity between two countries.  We focus only on the former, the national attributes of the guest country, because our aim is to capture the global view and thus does not necessarily assume the dyadic relationship led by a specific host country. 

A wide range of national attributes affecting news coverage is found across the studies~\cite{ahern1984determinants,ostgaard1965factors,ish1996us}
such as GNP per capita, territorial size, GDP, defense budget, population density, share in world trade, press freedom index, number of scientific publications, and Internet use.  We consider all these variables.  
We also include some variables such as world giving index, to consider the humanitarian view of the disasters. 


\section{The GDELT Project}

GDELT (Global Data on Events, Location, and Tone) is a recently developed event dataset containing more than 200 millions geolocated events with global coverage since 1979~\cite{leetaru2013gdelt}.  GDELT began with monitoring a wide range of international news sources, including AfricaNews, Agence France Presse, Associated Press Online, Associated Press Worldstream, BBC Monitoring, Christian Science Monitor, Facts on File, Foreign Broadcast Information Service, United Press International, and the Washington Post.  Now in cooperation with Google, it has expanded its sources to cover non-English news media.  Today it tracks news media in over 100 languages from the whole world.
The collected news articles are automatically categorized according to the CAMEO (Conflict and Mediation Event Observations) event coding taxonomy by using the open-source TABARI system\footnote{http://eventdata.parusanalytics.com/software.dir/tabari.html}.  
After the first release of GDELT, several studies have confirmed that the GDELT dataset performs as well or better than the previously widely-used datasets, such as ICEWS (Integrated Conflict Early Warning System) due to its large coverage and the improved automatic coding system~\cite{arva2013improving}.  


GDELT provides two types of datasets.  One is called the Event Database, coded by CAMEO taxonomy since 1979, and the other is the Global Knowledge Graph (GKG), an expanded dataset about `every person, organization, company, location, and over 230 themes and emotions from every news report' since 2013.  
We use the GKG dataset because it offers various fields to describe the characteristics of natural and man-made disasters, such as the type of a disaster, the number of news articles reporting the disaster, the number of victims, the location where the disaster occurred, and etc.  



\section{News geography of Disasters}

We first concentrate on news geography, the extent to which countries are represented in international disaster news.  
We show how different their news geographies are and then move on to examine the representativeness of each region for \textit{global} attention.  It directly relates with the external validity of previous studies about foreign news rooted on a single or several countries.  

We divide the world into seven regions according to the World Bank: East Asia \& Pacific, Europe \& Central Asia, Latin America \& Carribean, Middle East \& North Africa, North America, South Asia, and Sub-Saharan Africa.  The division mainly reflects geographical proximity.  
We map 10,009 news media into one of the seven regions according to the classification of Alexa, which is based on the nationality of website visitors.  A list of news media falling in each region becomes a basis to construct news geography seen by each region.  

We define the attention of a region, $r_i$, to a country, $c_j$, as the number of the disasters occurred in $c_j$ covered by news media of $r_i$.  We use the notation, $N_{r_i  \Longrightarrow c_j}$, for representing the attention of $r_i$ to $c_j$.  Then, we define news geography seen by $r_i$ as $\mathbf{N}_{r_i}=\{N_{r_i \Longrightarrow c_1}, N_{r_i \Longrightarrow c_2}, N_{r_i \Longrightarrow c_3}, ..., N_{r_i \Longrightarrow c_K}\}$ where $K$ is the number of the countries.

\begin{figure*} [hbt!]
  \begin{center}
    \begin{minipage}[b]{0.65\columnwidth}
      \begin{center}
      \includegraphics[width=\columnwidth]{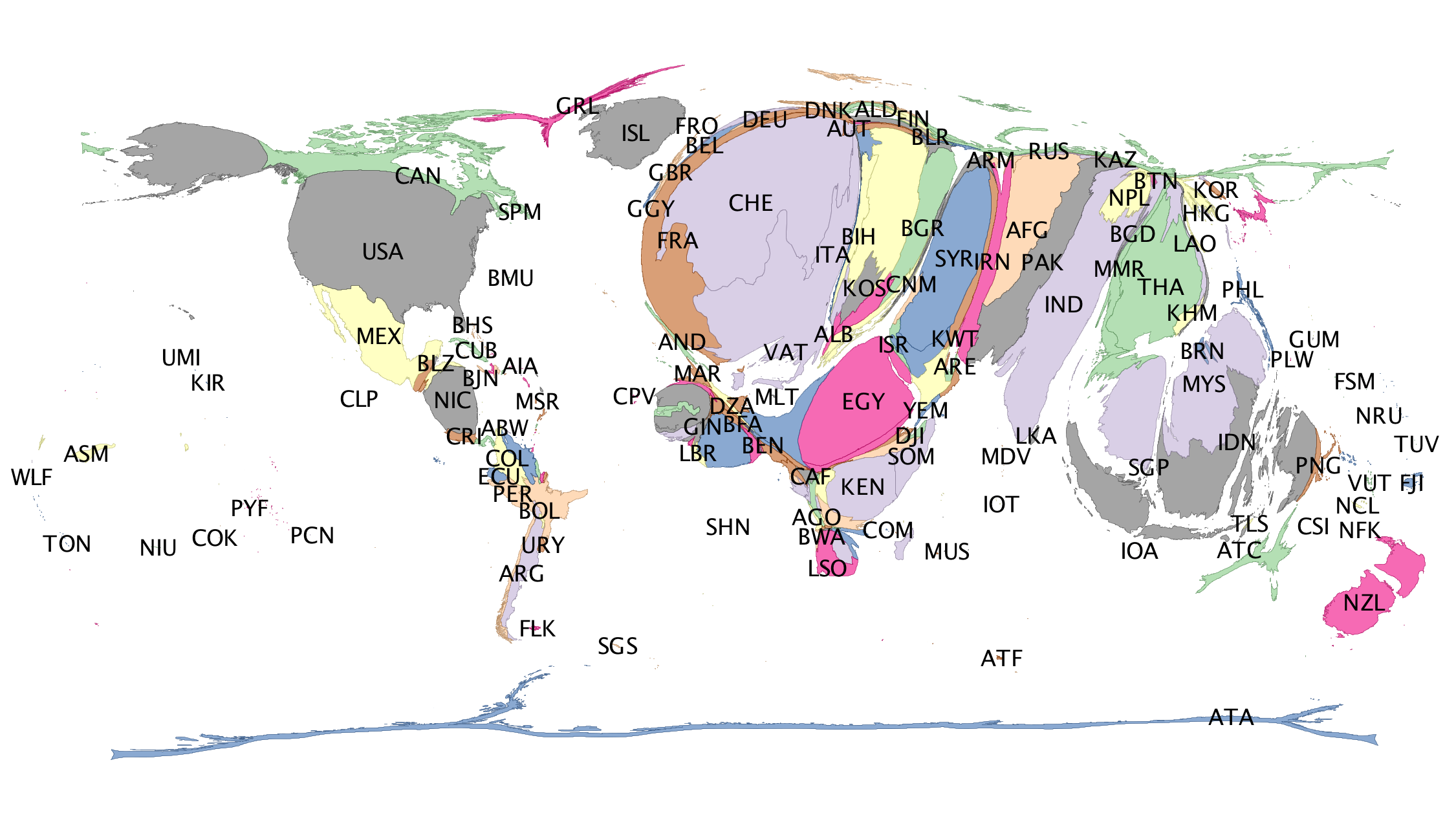}
      (a) East Asia \& Pacific
      \end{center}
    \end{minipage}
    \begin{minipage}[b]{0.65\columnwidth}
      \begin{center}
      \includegraphics[width=\columnwidth]{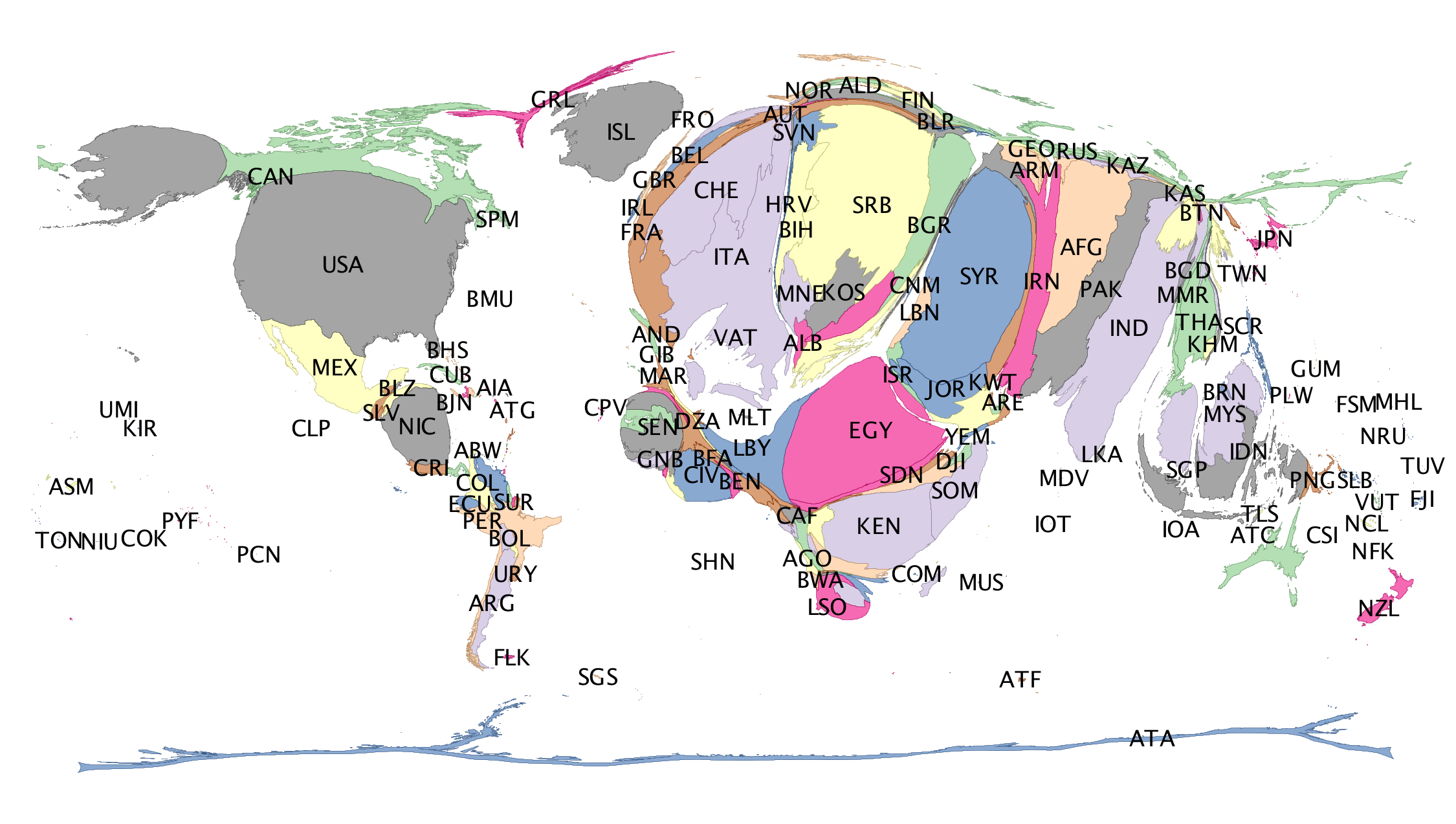}
      (b) Europe \& Central Asia
      \end{center}
    \end{minipage}
    \begin{minipage}[b]{0.65\columnwidth}
      \begin{center}
      \includegraphics[width=\columnwidth]{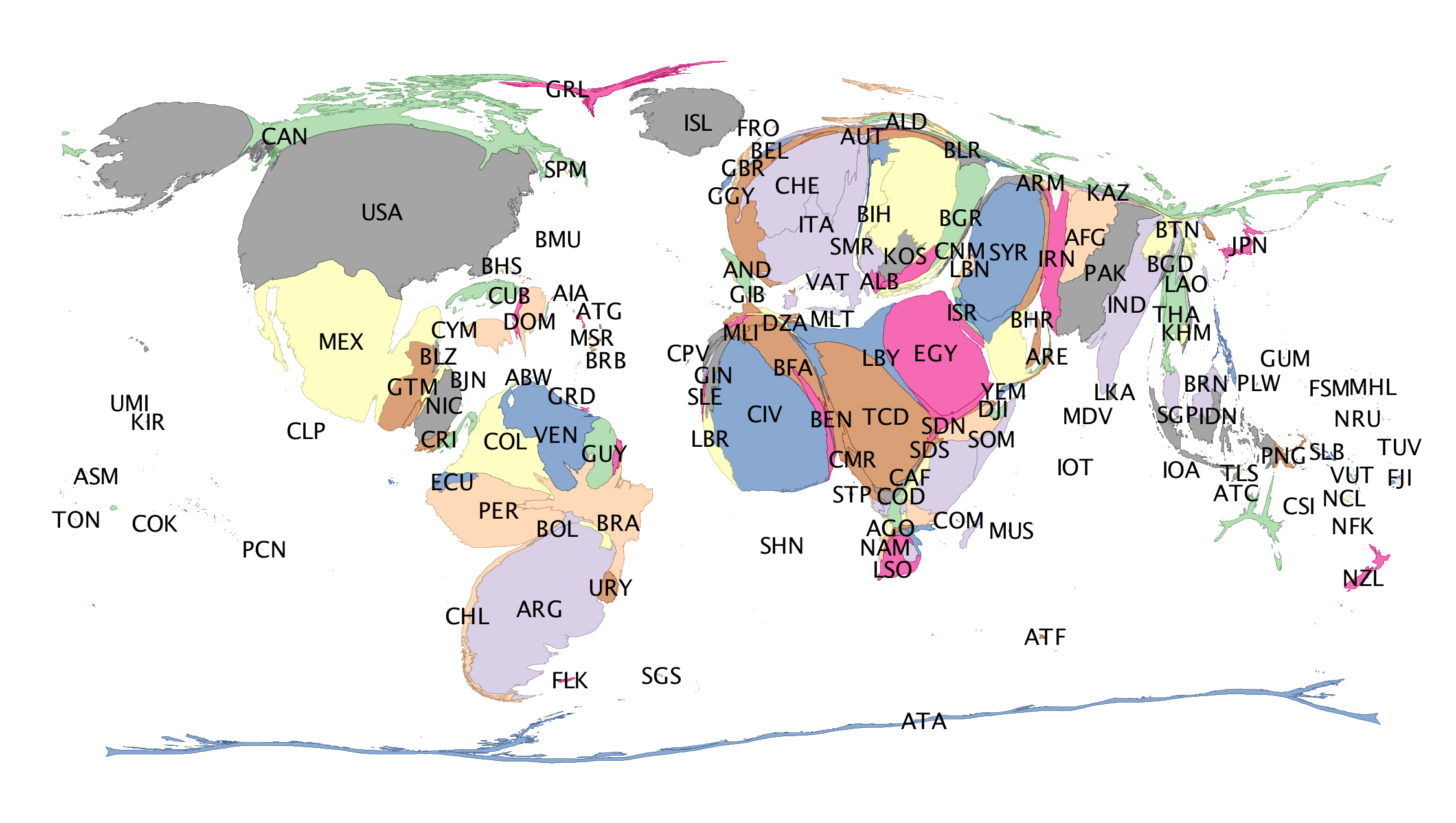}
      (c) Latin America \& Caribbean
      \end{center}
    \end{minipage}    
    \begin{minipage}[b]{0.65\columnwidth}
      \begin{center}
      \includegraphics[width=\columnwidth]{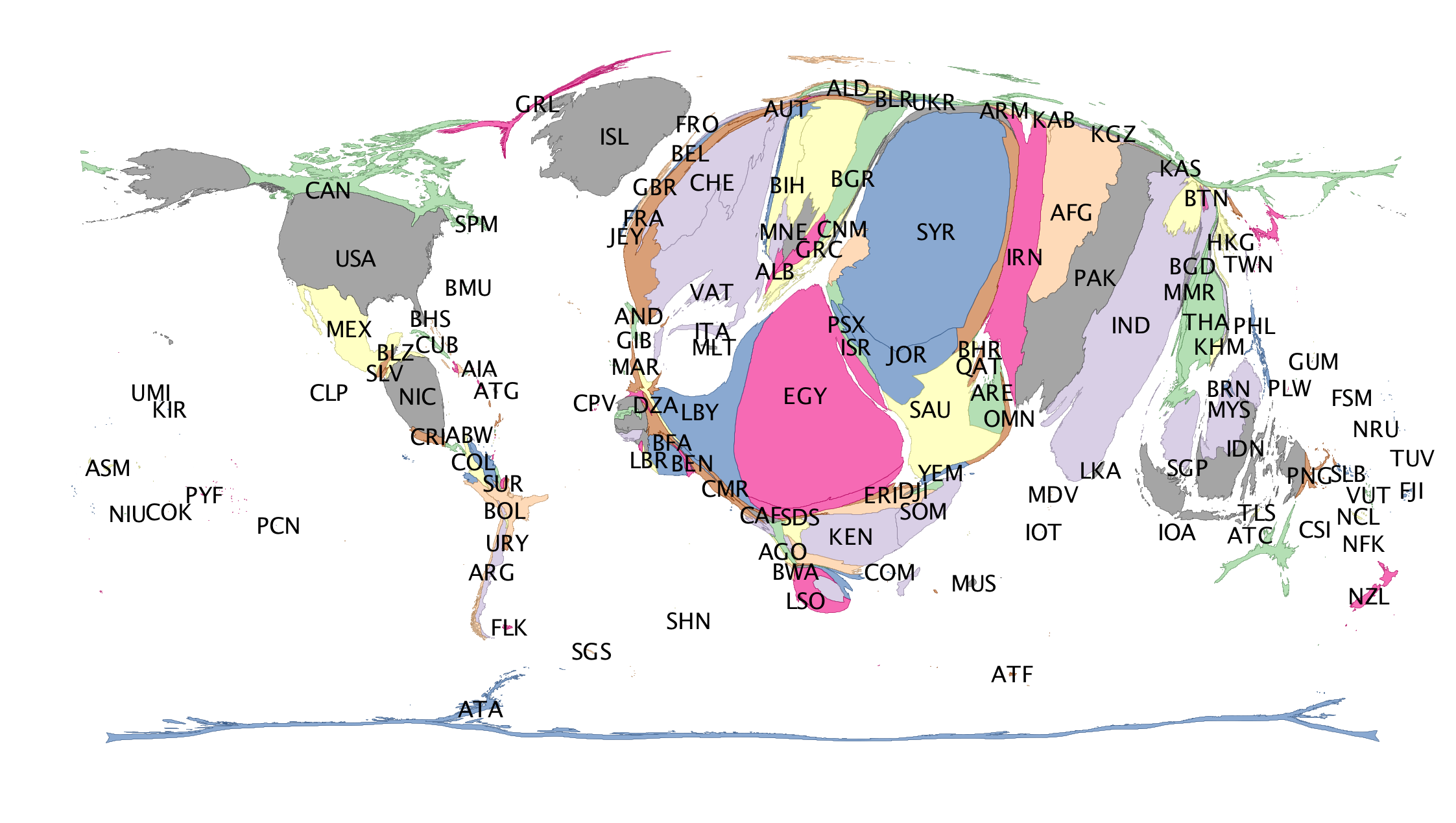}
      (d) Middle East \& North Africa
      \end{center}
    \end{minipage}     
    \begin{minipage}[b]{0.65\columnwidth}
      \begin{center}
      \includegraphics[width=\columnwidth]{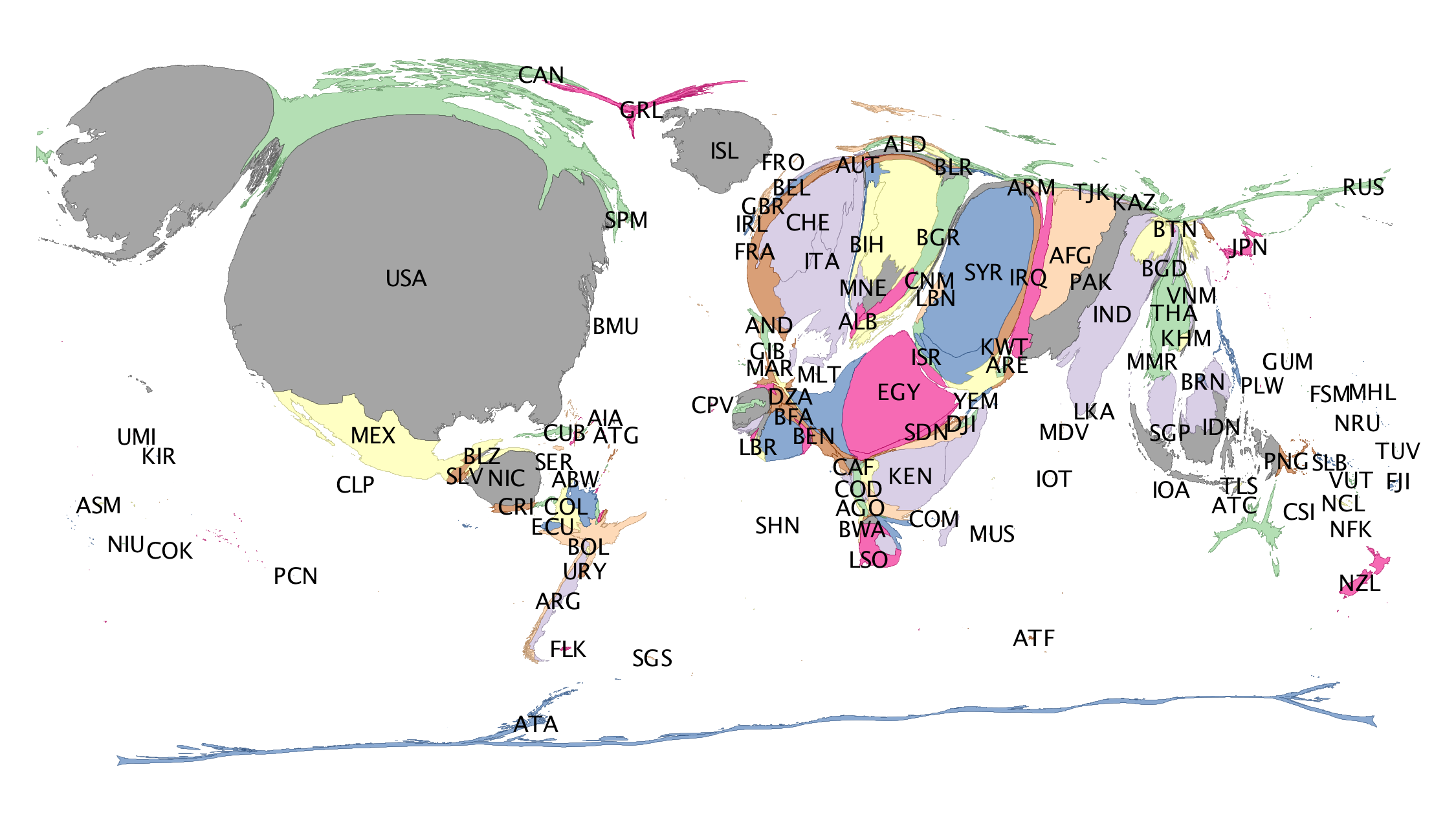}
      (e) North America
      \end{center}
    \end{minipage}
    \begin{minipage}[b]{0.65\columnwidth}
      \begin{center}
      \includegraphics[width=\columnwidth]{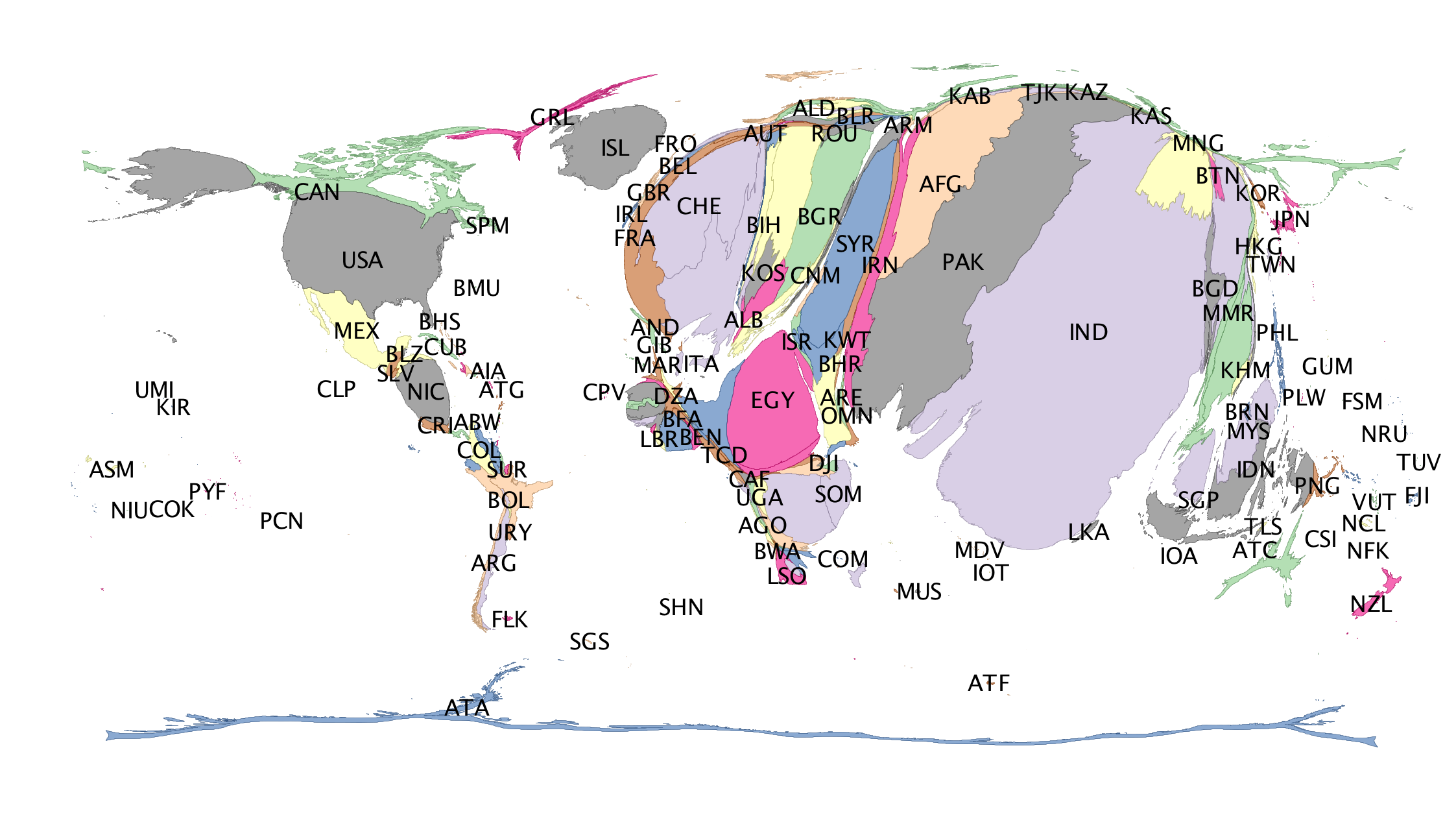}
      (f) South Asia
      \end{center}
    \end{minipage}
    \begin{minipage}[b]{0.65\columnwidth}
      \begin{center}
      \includegraphics[width=\columnwidth]{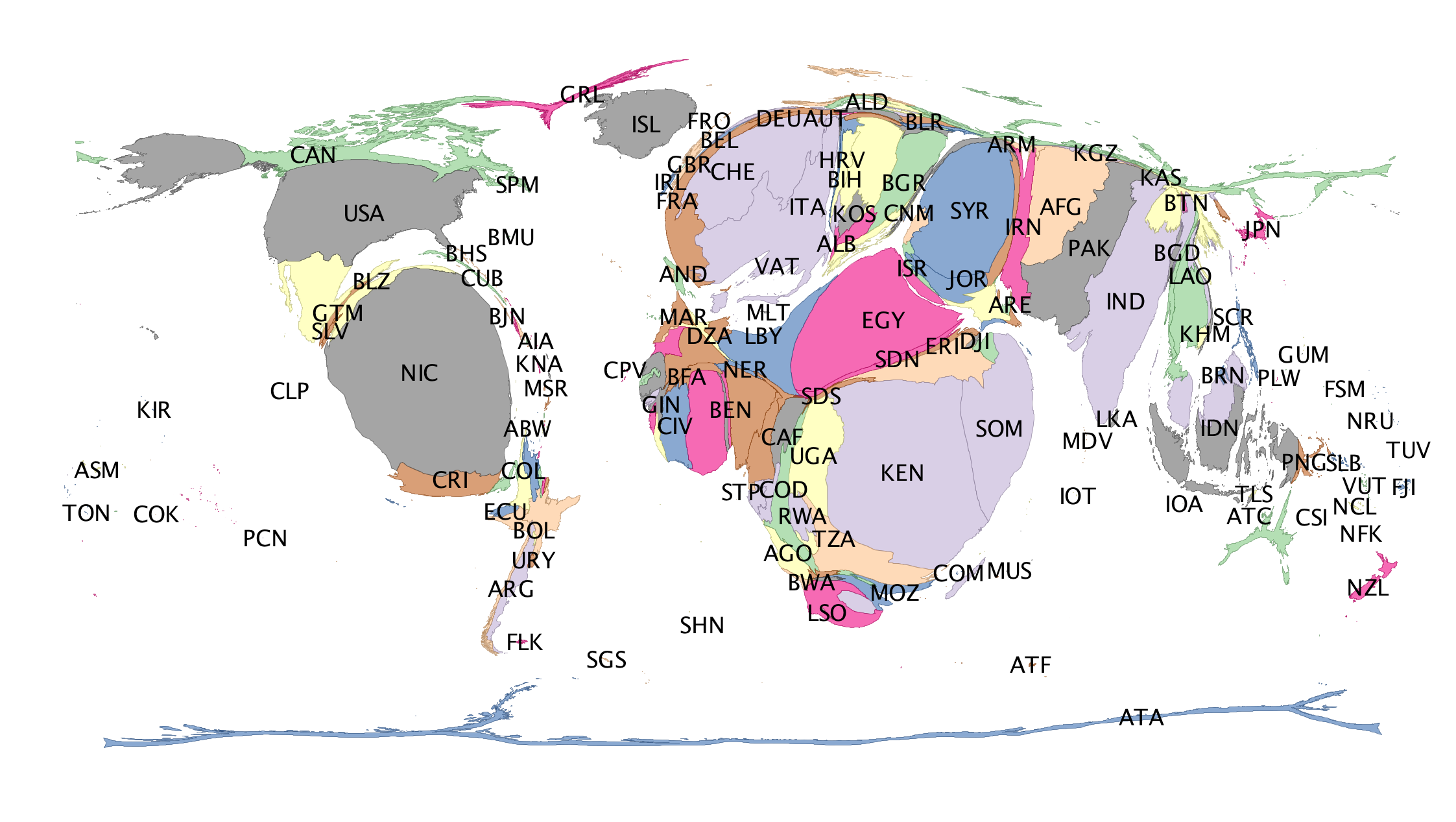}
      (g) Sub-Saharan Africa
      \end{center}
    \end{minipage}                    
  \caption{News geography seen by each region}  
  \label{fig:cartogram}
  \end{center}
\end{figure*}

Figure~\ref{fig:cartogram} shows the news geography seen by each region, as Cartogram, an intuitive visualization method of illustrating the territory of a country that is proportional to the assigned value.  In the figure, the size of a territory is proportional to $N_{r_i  \Longrightarrow c_j}$ in the news geography seen by $r_i$.  

By visual inspection, we observe clear differences of the news geography across the region.  Every region is overrepresented in the news geography seen by the corresponding region.  For example, disasters occurring in Latin America \& Caribbean are not frequently reported in other regions.  Similarly. Indonesia is well-recognized in $\mathbf{N}_{\textit{East Asia}}$, Serbia in $\mathbf{N}_{\textit{Europe}}$, Argentina in $\mathbf{N}_{\textit{Latin America}}$, Saudi Arabia in $\mathbf{N}_{\textit{Middle East}}$, USA in $\mathbf{N}_{\textit{North America}}$, India in $\mathbf{N}_{\textit{South Asia}}$, and Kenya in $\mathbf{N}_{\textit{Africa}}$.  This strong regionalism raises concerns about the external validity of studies of foreign news coverage based on a single country or region.  

At the same time, Figure~\ref{fig:cartogram} poses an interesting question about the over-representation of a certain country (e.g. Syria in news geography seen by North America) that cannot be explained by regionalism.  This relatively regionalism-free country could be explained by the proximity in another layer, such as politics, economy or culture, instead of geographic proximity.  Since the scope of this work is investigating global attention to disasters, rather than attention of a certain region or country, we do not study this further here.

Geographical divisions also make it feasible to quantify the representativeness of each region in reflecting global attention by comparing each region with the world in response to disasters.  
We define the attention of a region, $r_i$, to a disaster, $d_j$, as the number of news media in $r_i$ reporting $d_j$.  We use the notation, $A_{r_i \rightarrow d_j}$, to represent the attention of $r_i$ to $d_j$.  We then define the attention of $r_i$ to the whole set of disasters as $\mathbf{A}_{r_i}=\{A_{r_i \rightarrow d_1}, A_{r_i \rightarrow d_2},  ..., A_{r_i \rightarrow d_M}\}$, where $M$ is the number of the disasters.  For the sake of clarity, we note that previously we focus on the attention of a region to a country, but here we use the attention a region to a disaster. Then, the correlation between $\mathbf{A}_{r_i}$ and $\mathbf{A}_{\cup_j \{r_j\}}$ shows the representativeness of $r_i$ for global attention, where $\cup_j \{r_j\}=\{r_1, r_2, ..., r_7\}$ represents the world.

\begin{figure} [hbt!]
  \begin{center}
  \includegraphics[width=\columnwidth]{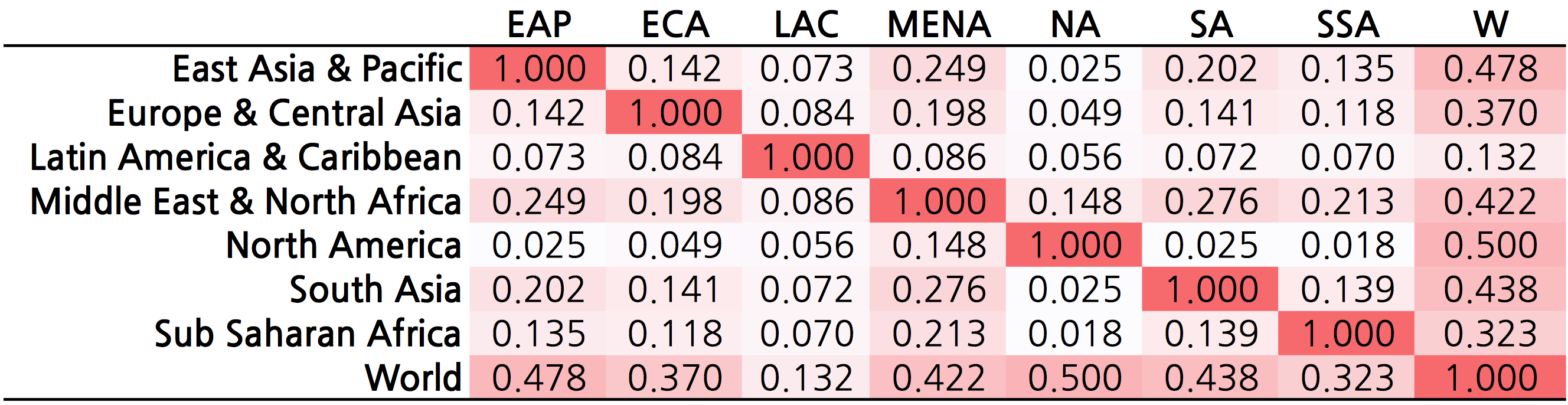}\vspace{-3mm}
  \caption{Correlation between $A_{r_i \rightarrow \cup_j {d_j}}$ and $A_{r_k \rightarrow \cup_j {d_j}}$ (All coefficients are statistically significant ($p$ $<$ .001))}  
  \label{fig:representativeness}
  \end{center}
\end{figure}

Figure~\ref{fig:representativeness} shows the representativeness of each region for global attention and the Spearman correlation between $\mathbf{A}_{r_i}$ and $\mathbf{A}_{r_j}$ (1 $<$ $i$, $j$ $<$ 7) as well.  The representativeness of each region is not high.  Although all the coefficients are positive, they are less than 0.5.  Particularly, we find $\mathbf{A}_{\textit{Latin America}}$ is quite different from global attention ($\rho$=0.132).  Strong regionalism, observed in news geography, is quantified.  Another interesting observation is that the correlation between a region and the world is higher than that between one region and another.  This implies that disasters covered by each region do not overlap with one another. This unique interest of each region makes it difficult for research relying on a single country or region to obtain external validity, but, on the other hand, it highlights the value of well-focused study.

\section{Determinants of global news coverage}

\subsection{Methods}

We build a hierarchical (mixed-effect) multiple regression model to examine what affects global news coverage of disasters.  We choose this model to control a random effect driven by variation rooted on country-level differences. Previous studies have shown that international news coverage varies significantly by country~\cite{wilke2012geography}. 

We define global news coverage of a disaster as the number of countries reporting the disaster, and use it as the dependent variable in our model.  
For 10,009 news media appearing in the GKG dataset, we extract the origin country of each news media from Alexa.
In our data, the range of global news coverage lies between 1 and 34. The median is 1 (mean: 1.78), indicating that a large fraction of disaster news is consumed within a single country. 

We consider 26 independent variables as candidates that might exert influence on the global news coverage of a disaster according to our theoretical orientations. The variables can be organized into three broad categories: (1) the attributes of a nation which measures political and economical status of a nation; (2) the attributes of a disaster; and (3) logistics of news gathering. 

We obtain national attributes for all the countries listed in the GDELT dataset from various sources, including the World Bank Open Data, which provides a wide range of up-to-date measures of 254 countries. Fifteen national variables are driven from it: GDP (gross domestic product) per capita, GNI (gross national income) per capita, military expenditure, population, land size, population density, merchandise exports (US\$), merchandise imports (US\$), number of scientific journal publications, unemployment rate, foreign aid received (US\$), Internet use (per 100 people), mobile cellular subscriptions (per 100 people), and homicide rate (per 100,000 people).  In addition, we create a trade variable as the sum of the magnitude of exports and imports.  While some information is not up-to-date, most of variables are reported for 2013.  

We additionally consider the index of press freedom~\cite{press.freedom}, the world giving index~\cite{giving.index}, and the political stability index~\cite{political.stability}.  

Index of press freedom~\cite{press.freedom} measures the number of violations of different kinds, including (a) the number of journalists who were jailed or killed in the connection with their activities; (b) the number of journalists abducted; (c) the number that fled into exile; (d) the number of physical attacks and arrests; (e) the number of media censored; and (f) violations of the right to information in foreign territory.  A possible score is from 0 (the best) to 100 (the worst). 

The world giving index (WGI)~\cite{giving.index} is compiled by the Charities Aid Foundation in 2012 based on data gathered by Gallup. They asked people which of the following three charitable acts they had undertaken in the past month: 1) donated money to an organization? 2) volunteered time to an organization? and 3)helped a stranger, or someone they did not know who needed help?  The proportion of the people answering yes is computed for each question, and the WGI is the mean of those three values for each country. 

Index of political stability and absence of violence/terrorism (denoted as political stability) measures perceptions on the likelihood that the government will be destabilized or overthrown by unconstitutional or violent means, including politically-motivated violence and terrorism. We collect the political stability of 215 countries from The Worldwide Governance Indicators (WGI) project~\cite{political.stability}.  The value scales from approximately -2.5 to 2.5, with higher values corresponding to better governance. For example, USA has 0.63, while Sudan has -2.27. 

We obtain attributes of disasters directly from the GKG dataset.
It marks a disaster as either man-made, natural, or both.  Also, a fine-grained subtype of a disaster, such as `Flooding' and `Landslide', is available.  We select one representative subtype for the disaster by considering the frequency of subtypes across all the disasters, while GDELT can assign multiple subtypes to one disaster.  For the matter of categorical coding, we select top 30 subtypes out of 274 subtypes, which account for 31.5\% of all disasters.  All the other types are coded as `other'. 
To measure unexpectedness of a disaster (denoted as UE\_disaster), we use the inverse of the frequency of the same-subtype disaster occurring in the corresponding country.  In other words, the more frequently a disaster occurs, the more expected the disaster is. 
We add variables about the impact of disasters: the number of people involved in the disaster, the type of people's involvement in the disaster (denoted as count type), and the type of people affected by the disaster (denoted as object type).  We also consider the country where the disaster occurs. 

We finally add one binary variable to show whether a disaster is reported by any of international news agencies, denoted as INAs covered.  This reflects logistics of news gathering, determining the news flow by gate-keeping.  We focus on three global news agencies, as many previous literature does: Agence France-Presse (AFP), Associated Press (AP) and Reuters. 

After considering the above variables, elimination of multicollinearity is a crucial step because multicollinearity distorts estimated coefficients of variables. 
We take three steps of analysis to select relevant variables.  First, we build a linear regression for each of the independent variables to see its predictive power for global news coverage of disasters.  In this step, we discard the homicide rate variable as it shows low significance in predicting the global coverage.  
Then, we compute the (Pearson) correlation coefficient between each pair of variables.  We find that a few national variables are correlated with each other (i.e., there are high positive correlations (where $r > $ 0.60) among GDP, GNI, GNI, Internet use, life expectancy, the number of scientific journal articles, political stability, and index of press freedom). We retain ``political stability'' as it is the most predictive factor among those eight variables.  By a variance inflation factor (VIF) test, we additionally remove the trade variable.  Lastly, through stepwise variable selection using AIC, we get the final regression model with 14 independent variables and an additional control variable (location).  We confirm no collinearity by a VIF test; all the remaining variables have VIF below than 5.3.  

For the analysis, we use the GKG dataset provided by the GDELT project that incorporates 3,574,627 events happening in 205 countries from April 2013 to July 2014. We extract 666,150 natural or man-made disasters and filter out disasters if any variable is missing.  As a result, we have 195,513 disasters to build the model. 



\subsection{Results}


We examine the explanatory power of variables in determining global news coverage of a disaster with our hierarchical multiple regression model. 
Seven national variables are entered in the first model.  Then, six disaster attributes are added to the second model for determining their unique contribution while  controlling for the national characteristics.  Finally, the full model is tested, including the simultaneous examination of all variables. We discuss only the contribution of individual variables in the full model due to lack of space.  We use a 0.05 level of statistical significance to evaluate the results of the regression analysis.


\begin{table}[t]
\scriptsize \frenchspacing
\begin{center}
\begin{tabular}{l D{)}{)}{11)2}@{} D{)}{)}{11)3}@{} D{)}{)}{11)3}@{} }
\toprule
                     & \multicolumn{1}{c}{Model 1} & \multicolumn{1}{c}{Model 2} &  \multicolumn{1}{c}{Full model} \\
\midrule
Intercept            & -0.87 \; (0.63)      & 0.33 \; (0.64)              & 0.19 \; (0.54)        \\
\hline
\hline
\textbf{\textsl{National attributes}} \\
\hline
Mobile subscriptions & 0.00 \; (0.00)^{**}  & 0.00 \; (0.00)^{**}   & 0.00 \; (0.00)^{**}   \\
log(Population)      & 0.08 \; (0.03)^{*}   & 0.07 \; (0.03)^{*}       & 0.06 \; (0.03)^{*}    \\
Political stability  & -0.17 \; (0.06)^{**} & -0.17 \; (0.05)^{**}  & -0.13 \; (0.04)^{**}  \\
\hline
\hline
\textbf{\textsl{Disaster attributes}} \\
\hline
Manmade disaster     &                      & -0.95 \; (0.02)^{***}                      & -0.76 \; (0.02)^{***} \\
Natural disaster     &                      & -1.06 \; (0.01)^{***}                      & -0.83 \; (0.01)^{***} \\
\# of affected people      &                      & 0.00 \; (0.00)^{***}                      & 0.00 \; (0.00)^{***}  \\
UE of disaster       &                      & -0.28 \; (0.07)^{***}                       & -0.17 \; (0.07)^{**}  \\
\textbf{Count type} \\
Kill            &                      & -0.33 \; (0.04)^{***}                       & -0.22 \; (0.04)^{***} \\
Other           &                      & -0.73 \; (0.04)^{***}                       & -0.52 \; (0.04)^{***} \\
Protest         &                      & -0.56 \; (0.04)^{***}                       & -0.45 \; (0.04)^{***} \\
Wound           &                      & -0.40 \; (0.04)^{***}                       & -0.26 \; (0.04)^{***} \\
\textbf{Object type} \\
Victims         &                      & -0.47 \; (0.22)^{*}                        & -0.44 \; (0.20)^{*}   \\
\textbf{Disaster subtype} \\
Radiation leak       &                      & 0.58 \; (0.37)                              & 0.65 \; (0.33)^{*}    \\
Toxic waste          &                      & 0.37 \; (0.17)^{*}                          & 0.06 \; (0.15)        \\
Aftershocks          &                      & 1.37 \; (0.37)^{***}                        & 1.08 \; (0.34)^{**}   \\
Flooding             &                      & 0.56 \; (0.14)^{***}                        & 0.33 \; (0.13)^{**}   \\
Heat wave            &                      & 0.33 \; (0.17)^{*}                          & 0.18 \; (0.15)        \\
Ice                  &                      & 0.58 \; (0.14)^{***}                        & 0.31 \; (0.13)^{*}    \\
Landslide            &                      & 0.67 \; (0.14)^{***}                        & 0.30 \; (0.13)^{*}    \\
Monsoon              &                      & 0.29 \; (0.15)^{*}                          & 0.14 \; (0.13)        \\
Severe weatehr       &                      & 1.10 \; (0.17)^{***}                       & 0.07 \; (0.15)        \\
Tsunami              &                      & 0.33 \; (0.15)^{*}                         & 0.20 \; (0.13)        \\
Other                &                      & 0.38 \; (0.14)^{**}                        & 0.20 \; (0.13)        \\
\hline
\hline
\multicolumn{4}{l}{\textbf{\textsl{Logistics of news gathering}}}\\
\hline
INAs covered          &                      &                       & 3.74 \; (0.02)^{***}  \\
\midrule
$R^2$                & 0.03090539           & 0.07369887                  & 0.2536145              \\             
AIC                  & 802417.49            & 793681.70             & 750501.26             \\
BIC                  & 802519.33            & 794241.77            & 751071.51             \\
Log Likelihood       & -401198.75           & -396785.85            & -375194.63            \\
Num. groups          & 74                   & 74                    & 74                    \\
\bottomrule

\multicolumn{4}{l}{\scriptsize{$^{***}p<0.001$, $^{**}p<0.01$, $^*p<0.05$}}
\end{tabular}
\caption{Hierarchical multiple regression predicting global news coverage (N = 195,513). }
\label{table:coefficients}
\end{center}
\end{table}

Table~\ref{table:coefficients} shows the regression result with estimated coefficient and its statistical significance.  In the first model, three national variables, which are log(population), mobile subscription, political stability, have a significant effect on the dependent variable and explain 3.1\% of its variance. The figures for the second model, in which disaster variables are added to the first model, indicate that the characteristics of a disaster themselves explain an additional 4.3\% of the variance.  Together, the national and disaster variables explain 7.9\% of the variance in global news coverage.  
 
In the final step, the newly added variable, INAs covered, explains an additional 18\% of the variance, resulting in a total of 25.4\% of the variance in global news coverage.  Its gain is much more than the amount of variance explained by the national and disaster variables.  The explanatory power of our model is comparable with previous studies in this research area; in the study of finding systemic determinants of news coverage of 38 countries~\cite{wu2000systemic}, 20 out of 38 models have lower $R^2$ than ours. 





\subsubsection{National attributes}  We find that out of 7 variables considered, only two are statistically significant. Population has a positive coefficient, but its effect is marginal.  High population of a country commonly leads to more emigrants to other countries.  For example, China and India are major nationalities of US immigrants, in spite of their remoteness to US.  As Lacy et al. point out, news coverage of newspapers is influenced by audience demand~\cite{lacy1989impact}.  Higher number of immigrants possibly explain more news coverage for them.  This relationship would be clarified when a guest, a host country framework for foreign news coverage is adopted, when the number of immigrants can be counted. 

The political stability is negatively correlated with global news coverage.  In other words, disasters happening in politically unstable countries receive more global attention.  This finding is aligned with the study by Masmoudi that reveals the Western news media intentionally cover crisis or conflicts of  unstable countries so that the stereotype of them can be reinforced~\cite{masmoudi1979new}. Even though we find the possibility of such a tendency from the aggregated news media, we do not attempt to quantify the contribution of Western news media to this.  In-depth analysis is required to assess the universality of reinforcing stereotype.  

\subsubsection{Disaster attributes} 
The negative coefficients of man-made disaster and natural disaster mean that a disaster tagged as both man-made and natural disasters is more likely to get global attention than when only one theme is tagged.  It implies that when a natural disaster happens, a complex situation where human factors are involved is likely to be covered by news media.

We find that the number of affected people is statistically significant with a marginal effect. This supports the previous finding that the number of killed people is an important factor for the natural disaster news coverage~\cite{gaddy1986earthquake}. In our study, not only killed people but also affected people have been taken into account and we find that it has significant explanatory power.  

The number of affected people greatly varies across disasters.  For example, sometimes a few tens of thousands of people are evacuated when serious flooding occurs. 
Given that the significance of the number of killed people is still debatable~\cite{adams1986whose}, we believe that follow-up investigations in various settings are vital for assessing the significance of the affected people in global news coverage.  
    
Unexpectedness of disaster subtype is negatively correlated with global news coverage. This finding counters what had been found in the theory of newsworthiness~\cite{galtung1965structure}.  We explain a possible mechanism in the discussion section.  

The count type is a categorical variable. The beta coefficients of category show the relative predictive power of each type compared to the base type, which is Kidnap in our model. 
Among the count types we consider, we find that the Kidnap type tends to get the most global attention compared to other type: Kill (-0.22), Wound (-0.26), Protest (-0.45), Other (-0.52). The negative coefficient of the Other type means that fewer countries report the Other type than they report the Kidnap type.

Lastly, some disaster types are more favorable globally than others. Although marginal, five disaster types are statistically significant and positively related to the number of countries covering the disaster: Aftershocks (1.08), Radiation leak (0.65), Flooding (0.33), Ice(0.31), and Landslide(0.30).

\subsubsection{Logistic of news gathering}  INAs covered has the largest beta coefficients (3.74). Its positive sign indicates that a disaster covered by INA is more likely to get global attention. This is expected; however, the extent of the effect is not expected.  INAs covered alone explains 18\% of the variance in global news coverage.  
This result is in line with previous work reporting that the presence of INAs is a primary predictor of the amount of news coverage about the country~\cite{wu2000systemic}.  We also agree with Wu's argument that the most news media sites are dependents of INAs because the cost of managing correspondents to investigate foreign issues is higher than using news copy of INAs. 
We discover that the INAs still play a prominent role in expanding news coverage of a foreign disaster.

\balance

\section{Discussion}

Since our work largely depends on the GDELT dataset, addressing possible concerns about the completeness of the GDELT dataset is crucial. 

First, GDELT archives news articles every day.  If an event lasts more than a day, then it is counted as multiple events.  
We suppose that this affects the significance of unexpectedness of the disaster in our model.  
We define unexpectedness of a disaster as the inverse of the frequency of the same-category disaster occurred in the corresponding country.  For example, flooding in Malaysia or Indonesia is not as newsworthy as that in Saudi Arabia.   
However, contrary to our expectation and to previous findings, the unexpectedness of a disaster shows a negative coefficient in our model.  
Continuity of an event is suggested as one of the factors that influence newsworthiness of the event~\cite{galtung1965structure}.  Events that happen over multiple days reinforce their newsworthiness over time and are likely to get more attention than one-time events.  As a result, unexpectedness of a event that lasts for a few days and gets high attention is penalized due to the daily-basis archiving scheme in GDELT. Tracking the same event across days in GDELT is essential to resolve this issue, and it would be helpful for future work that needs correct temporal information of events.   

Another concern of using GDELT to study the global news coverage is its representativeness. 
A lot of US media are tracked by GDELT, as the strongest player in the media industry.  Thus, the number of news articles about a disaster is readily influenced by US news media. 
To neutralize this possible bias, we define global attention to a disaster as the number of countries covering the disaster.  
This equalizes the contribution of news media in each country to global attention.  While this simplifies the international power relationship, it also captures well the news geography and news flow among countries, which is essential to our research question.  

Lastly, the recall and the accuracy of the GDELT dataset in extracting disasters is vital. 
Although it is hard to measure the completeness of the GDELT project mainly due to the lack of the large-scale ground-truth, many studies with the GDELT dataset have confirmed that the GDELT project is reliable~\cite{arva2013improving}.  
We also manually check the recall of the significant disasters occurring in 2014~\cite{disaster@wikimedia}.  
Among the disasters listed in Wikipedia from January to May 2014, all the worst 30 disasters appear correctly in the GDELT dataset.

\bibliographystyle{abbrv}
\bibliography{cj_symposium_main}  

\end{document}